\documentclass[aps,reprint,showpacs,showkeys,preprintnumbers,groupedaddress,amsmath,amssymb,showkeys]{revtex4-1}

\usepackage{graphicx}% Include figure files
\usepackage{dcolumn}% Align table columns on decimal point
\usepackage{bm}% bold math
\usepackage{hyperref}% add hypertext capabilities
\usepackage[dvips]{color}

\begin{document}

% Use the \preprint command to place your local institutional report
% number in the upper righthand corner of the title page in preprint mode.
% Multiple \preprint commands are allowed.
% Use the 'preprintnumbers' class option to override journal defaults
% to display numbers if necessary

\preprint{}

%Title of paper

\title{Hopping conductance and macroscopic quantum tunneling effect in three dimensional Pb$_x$(SiO$_2$)$_{1-x}$ nanogranular films}

% repeat the \author .. \affiliation  etc. as needed
% \email, \thanks, \homepage, \altaffiliation all apply to the current
% author. Explanatory text should go in the []'s, actual e-mail
% address or url should go in the {}'s for \email and \homepage.
% Please use the appropriate macro foreach each type of information

% \affiliation command applies to all authors since the last
% \affiliation command. The \affiliation command should follow the
% other information
% \affiliation can be followed by \email, \homepage, \thanks as well.
\author{Xiu-Zhi Duan}
\affiliation{Tianjin Key Laboratory of Low Dimensional Materials Physics and
Preparing Technology, Department of Physics, Tianjin University, Tianjin 300354,
China}
\author{Zhi-Hao He}
\affiliation{Tianjin Key Laboratory of Low Dimensional Materials Physics and
Preparing Technology, Department of Physics, Tianjin University, Tianjin 300354,
China}
\author{Yang Yang}
\affiliation{Tianjin Key Laboratory of Low Dimensional Materials Physics and
Preparing Technology, Department of Physics, Tianjin University, Tianjin 300354,
China}
\author{Zhi-Qing Li}
\email[Corresponding author, e-mail: ]{zhiqingli@tju.edu.cn}
\affiliation{Tianjin Key Laboratory of Low Dimensional Materials Physics and
Preparing Technology, Department of Physics, Tianjin University, Tianjin 300354,
China}

%\homepage[]{Your web page}
%\thanks{}
%\altaffiliation{}

%Collaboration name if desired (requires use of superscriptaddress
%option in \documentclass). \noaffiliation is required (may also be
%used with the \author command).
%\collaboration can be followed by \email, \homepage, \thanks as well.
%\collaboration{}
%\noaffiliation

\date{\today}

\begin{abstract}
We have studied the low-temperature electrical transport properties of Pb$_x$(SiO$_2$)$_{1-x}$ ($x$ being the Pb volume fraction) nanogranular films with thicknesses of $\sim$1000 nm and $x$ spanning the dielectric, transitional, and metallic regions. It is found that the percolation threshold $x_c$ lies between 0.57 and 0.60. For films with $x$$\lesssim$0.50, the resistivities $\rho$ as functions of temperature $T$ obey $\rho\propto\exp(\Delta/k_BT)$ relation ($\Delta$ being the local superconducting gap and the $k_B$ Boltzmann constant) below the superconducting transition temperature $T_c$ ($\sim$7\,K) of Pb granules. The value of the gap obtained via this expression is almost identical to that by single electron tunneling spectra measurement. The magnetoresistance is negative below $T_c$ and its absolute value is far larger than that above $T_c$ at a certain field. These observations indicate that single electron hopping (or tunneling), rather than Cooper pair hopping (or tunneling) governs the transport processes below $T_c$. The temperature dependence of resistivities shows reentrant behavior for the 0.50$<$$x$$<$0.57 films. It is found that single electron hopping (or tunneling) also dominates the low-temperature transport process for these films. The reduction of the single electron concentration leads to an enhancement of the resisivity at sufficiently low temperature. For the 0.60$\lesssim$$x$$\lesssim$0.72 films, the resistivities sharply decrease with decreasing temperature just below $T_c$, and then show dissipation effect with further decreasing temperature. Treating the conducting paths composed of Pb particles as nanowires, we have found that the $R(T)$ data below $T_c$ can be well explained by a model that includes both thermally activated phase slips and quantum phase slips.
\end{abstract}

% insert suggested PACS numbers in braces on next line
% insert suggested keywords - APS authors don't need to do this
%\keywords{ Transport properties, Granular composites}

%\maketitle must follow title, authors, abstract, \pacs, and \keywords
\maketitle

% body of paper here - Use proper section commands
% References should be done using the \cite, \ref, and \label commands
%\section{}
% Put \label in argument of \section for cross-referencing
%\section{Introduction\label{sec1}}
%\subsection{}
%\subsubsection{}

\section{Introduction}
The transport properties near the superconductor-insulator transition (SIT) in disordered systems continue to attract intense theoretical and experimental interest~\cite{no1, no2, no3, no4, no5, no6, no7, no8, no9, no10, no11, no12, no13, no14, no15, no16, no17, no18, no19, no20, no21,no22, no23, no24}. In this regards, significant attention has been paid to two limiting disordered material systems, the uniform and granular systems. Here the uniform cases are referred to the systems with a potential inhomogeneity only on an atomic scale, while the materials with inhomogeneities that substantially exceed atomic dimensions are called granular systems. For the uniform cases, extensive investigations have been carried out on two dimensional (2D) systems and a wealth of unusual and striking phenomena, including the quantum dissipation in the vortex state~\cite{no10}, the quantum metallic state~\cite{no20, no21}, Berezinskii-Kosterlitz-Thouless (BKT) transition~\cite{no17}, and the universal scaling relation for sheet resistance with magnetic field and temperature~\cite{no10, no25}, have been revealed. For the granular systems, early studies near the SIT were focused on the origin of critical sheet resistance $R_c = h/(2e)^2$ (where $h$ is the Planck constant and $e$ is the electronic charge) for the onset of superconductivity in 2D films~\cite{no1, no2, no3, no4, no5, no6, no7}. The renewed interest was partially spurred by the theoretical suggestion of disorder-induced spatial inhomogeneity in the superconducting order parameters~\cite{no26}. Noted that the high-temperature superconductors are intrinsically disordered, which could lead to self-induced superconducting droplets and render the originally uniform system to become nonuniform~\cite{no27, no28}.

For the granular superconductors, the previous results on the temperature dependent behaviors of the resistivity below the superconducting transition temperature $T_c$ are quite inconsistent. In discontinuous aluminum~\cite{no29}, Al-Al$_2$O$_3$~\cite{no30}, and Al-Ge~\cite{no8} films, it has been reported that below $T_c$ the resistivity increases more rapidly with decreasing temperature than that above $T_c$ and the logarithm of the resistivity $\log\rho$ (or $\ln\rho$) increases linearly with $T^{1/2}$.  While in Pb~\cite{no29}, Ga or Al~\cite{no31}, and Bi~\cite{no31, no32} films,  it has been found that the resistivity increases exponentially with decreasing temperature below $T_c$, i.e., $\ln(\rho/\rho_0)\propto1/T$ with $\rho_0$ being a constant. In addition, the recent theory~\cite{no33} predicates that the electron hops via inelastic cotunneling mechanism at $T_1$$\lesssim$$T$$\lesssim$$T_c$ in the insulating phase, where $T_1 \approx 0.1\sqrt{E_c\delta}$ is a character temperature, $E_c$ and $\delta$ are the charging energy and the mean energy level spacing in a single grain, respectively. The inelastic cotunneling process would lead to an activation form resistivity. Thus the  transport properties of the granular superconductor in the insulating regime need further investigation. On the other hand, the dissipation effect has been observed in the quasi one dimensional (1D) superconducting nanowires~\cite{no34, no35, no36, no37, no38, no39} and 2D disordered superconductors~\cite{no10,no17} near the SIT. Thus to check whether the dissipation effect exists in three dimensional (3D) granular superconductor is interesting and nontrivial.

Considering Pb is immiscible with SiO$_2$ and the superconducting transition temperature of granular Pb films is close to that of the bulk metal~\cite{no40}, we systematically investigate the electrical transport properties of Pb$_x$(SiO$_2$)$_{1-x}$ nanogranular films, where $x$ is the volume fraction of Pb. For the insulating films, we found that the temperature dependence of resistivity obeys the activation form hopping law from $\sim$7\,K to $T_s$ ($T_s$ is the temperature below which the resistivity deviates to the activation form law and tends to be saturated, the value of $T_s$ is sample dependent and varies from $\sim$3 to $\sim$4\,K for our films), and gradually tends to be saturated with further decreasing temperature below $T_s$. The dissipation effect is observed in the films with $x$ slightly greater than $x_c$, where $x_c$ is percolation threshold above which the films show metallic characteristic in transport properties. We report our interesting observations in the following discussions.

\section{Experimental Method}
Our Pb$_x$(SiO$_2$)$_{1-x}$ films were deposited at room temperature by co-sputtering Pb and SiO$_2$ targets in Ar atmosphere. The details
of the deposition procedures were described previously~\cite{no40}. The films were simultaneously deposited on the the glass (Fisherfinest premium microscope slides) and polyimide (Kapton) substrates for the transport and composition (polyimide) measurements. The thicknesses of the films ($\sim$1000\,nm) were measured using a surface profiler (Dektak, 6M). The Pb volume fraction $x$ in each film was obtained from the energy-dispersive X-ray spectroscopy
analysis. The microstructure of the films was characterized by transmission electron microscopy (TEM, Tecnai G2 F20). The resistivities variation with temperature, as well as variation with magnetic field, were measured using standard four-probe method. The temperature and magnetic field environments were provided by a physical property measurement system (PPMS-6000, Quantum Design). For films with large $x$ (small resistivity), both the current source and the voltmeter were provided by the model 6000 PPMS controller. While for the small $x$ (high resistivity) films, a Keithley 236 and a Keithley 2182A  were used as current source and voltmeter, respectively. The narrow rectangle shape films (2\,mm$\times$10\,nm), defined by mechanical masks, were used for transport measurement. To obtain good contact, the Ti/Au electrodes were deposited on the films.

\begin{figure}
\begin{center}
\includegraphics[scale=0.9]{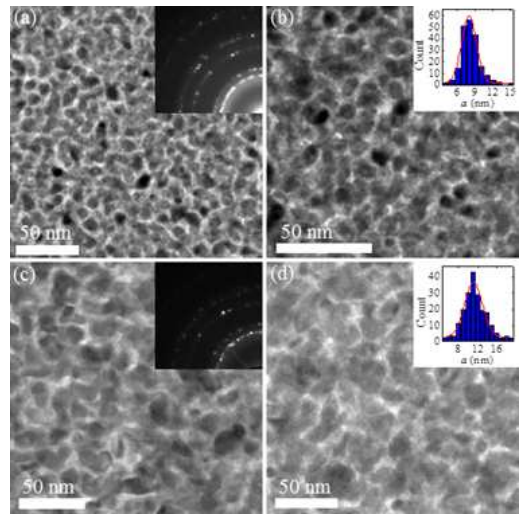}
\caption{Bright-field TEM images for Pb$_x$(SiO$_2$)$_{1-x}$ films with $x$ values of (a) 0.47, (b) 0.50, (c) 0.60, and (d) 0.65. The insets in (a) and (c) are the selected-area electron-diffraction patterns of corresponding films, and the insets in (b) and (d) show the corresponding grain size distribution histograms.}\label{LiTEM}
\end{center}
\end{figure}

\section{Results and Discussions}
Figure~\ref{LiTEM} shows bright-field TEM images of four representative films with $x$$=$$0.47$, 0.50, 0.60, and 0.65. The bright and dark regions in each image are SiO$_2$ and Pb, respectively. Only the diffraction corresponding to face-centered cubic Pb can be observed in the SAED pattern (the inset in Fig.~\ref{LiTEM}), indicating that SiO$_2$ is amorphous. The mean-size of Pb grains, $a$, obtained by taking into account $\sim$200 grains for each film, increases with increasing $x$.  It is found that mean-sizes of Pb grains are less than $\sim$8\,nm for those $0.47$$\lesssim$$x$$\lesssim$0.50 films, and about $\sim$9\,nm for the $0.51$$\lesssim$$x$$\lesssim$0.60 films, then vary between $\sim$11 and $\sim$17\,nm for the 0.65$\lesssim$$x$$\lesssim$0.74 films.

\begin{figure}
\begin{center}
\includegraphics[scale=0.80]{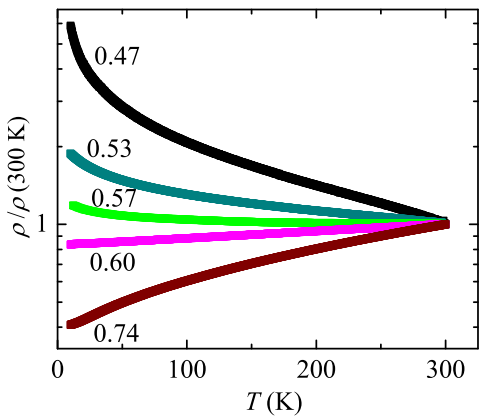}
\caption{Normalized resistivity as a function of temperature from 300 down to 10 K for the films with  $x$$=$$0.47,\, 0.53,\, 0.57,\, 0.60$, and 0.74.}\label{LiRT10to300}
\end{center}
\end{figure}

Figure~\ref{LiRT10to300} shows temperature dependence of normalized resistivity from 300 down to 10\,K for some representative films, as indicated. The resistivities of the $x$$\lesssim$0.57 films increase with decreasing temperature in this temperature range. While the resistivity decreases with decreasing temperature for the $x$$\gtrsim$0.60 films. Thus the percolation threshold $x_c$ for the metal-insulator transition lies between 0.57 and 0.60, which is identical to that reported in Ref.~\onlinecite{no40}. Next, we discuss the low temperature electrical transport properties of the films with $x$$<$$x_c$ and $x$$\gtrsim$$x_c$ in detail.

\begin{table}
\caption{\label{TableI} Relevant parameters for some Pb$_x$(SiO$_2$)$_{1-x}$ films. Here $x$ is volume fraction of Pb,  $\rho_{01}$ and $T_0$ are the parameters in Eq.~(\ref{EqESVRH}),  $\rho_{02}$ and $\Delta_{01}$ are the parameters in Eq.~(\ref{EqActiavtion}), $\Delta_{02}$ is the deduced zero temperature superconducting gap by considering the temperature effect of $\Delta$ in Eq.~(\ref{EqActiavtion}).}
\begin{ruledtabular}
\begin{center}
\begin{tabular}{ccccccc}%\hline \hline
$x$    & $\rho$(300\,K)  &  $\rho_{01}$    & $T_0$   &  $\rho_{02}$      &  $\Delta_{01}$   &   $\Delta_{02}$ \\
       & ($\Omega$\,cm) & ($10^2$\,$\Omega$\,cm) & (K)     &   ($\Omega$\,cm) & ($10^{-22}$\,J)  &  ($10^{-22}$\,J)\\  \hline
0.47   & 3.29$\times$10$^2$     & 5.76  &  14.94  & 222                  &    2.42    &   1.86  \\
0.48   & 2.94$\times$10$^2$     & 4.00  &  3.50    &   58.0              &    2.38    &   1.59  \\
0.49   & 2.04$\times$10$^2$     & 2.34  &  4.22   &   32.1               &    2.40    &   1.63  \\
0.50   & 67.0                   &        0.79         &  3.00   &   10.9              &    2.41    &   1.52  \\
0.53   & 1.11                   &          $-$        &  $-$    &   $-$               &     $-$    &  $-$    \\
0.57   & 1.03                   &          $-$        &  $-$    &   $-$               &     $-$    &  $-$    \\
0.60   & 1.22$\times$10$^{-1}$  &          $-$        &  $-$    &   $-$               &     $-$    &  $-$    \\
0.74   & 1.68$\times$10$^{-3}$  &          $-$        &  $-$    &   $-$               &     $-$    &  $-$    \\

\end{tabular}
\end{center}
\end{ruledtabular}
\end{table}

\subsection{Hopping conductance in the insulating regime}
\begin{figure}
\begin{center}
\includegraphics[scale=1]{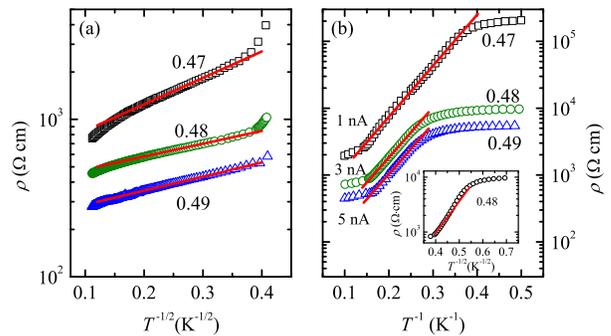}
\caption{(a) Logarithm of resistivity as a function of $T^{-1/2}$ from 40 down to 10 K for the $x$$=$$0.47,\, 0.48$, and 0.49 films. The symbols are the experimental data and the straight solid lines are least-squares fits to Eq.~(\ref{EqESVRH}). (b) Logarithm of resistivity as a function of the reciprocal of temperature from $\sim$10 down to 2\,K for the same films in (a). The symbols are the experimental data and the straight solid lines are least-squares fits to  Eq.~(\ref{EqActiavtion}). The inset represents the logarithm of resistivity as a function of $T^{-1/2}$ from $\sim$8 to 2\,K for the $x=0.48$ film. }\label{LiRTLTLX}
\end{center}
\end{figure}

Figure~\ref{LiRTLTLX}(a) shows the resistivity as a function of $T^{-1/2}$ in single logarithmic scale for three $x$$<$$0.50$ films, as indicated. From $\sim$40\ down to $\sim$8\,K, the $\log\rho$ (or $\ln \rho$) of each film varies linearly  with $T^{-1/2}$, i.e.,
\begin{equation}\label{EqESVRH}
 \rho=\rho_{01} \exp \left(\frac{T_0}{T}\right)^{1/2},
\end{equation}
with  $\rho_{01}$ being a constant independent of temperature and $T_0$ a material-dependent constant. %then increases more sharply below $\sim$7\,K.
The behavior of resistivity described by Eq.~(\ref{EqESVRH})  often appears in the disordered semiconductor~\cite{no41,no42, no43,no44,no45} and insulating regime of granular metals~\cite{no46, no47}. In disordered semiconductor, the Coulomb interactions between the charge carriers open a Coulomb gap near the Fermi level, which causes the Mott-type variable-range-hopping (VRH) conduction process cross over to the Efros-Shklovskii-type VRH process when the temperature is sufficient low~\cite{no41, no42}. Thus the transport process determined by Eq.~(\ref{EqESVRH}) in disordered semiconductor is called  Efros-Shklovskii-type VRH conduction. In granular metals, Sheng and coworkers~\cite{no46, no47} and Wu \emph{et al}~\cite{no48} have considered the nonuniformity of the metallic granule size and analyzed the conduction process of thermal activated charge carriers through the percolation path. Then Eq.~(\ref{EqESVRH}) could be obtained. Recently, several researchers coming from different groups have reconsidered the origin of Eq.~(\ref{EqESVRH}) in the insulator phase of granular metals~\cite{no49, no50, no51}. They have suggested that the `soft gap' being similar to the Coulomb gap still exist in the granular metals, and the electron conduction can occur through `cotunneling' of electrons between distant metallic granules
via a chain of intermediate virtual states. The solid lines in Fig.~\ref{LiRTLTLX}(a) are least-squares fits to Eq.~(\ref{EqESVRH}). The value of the adjustable parameter $T_0$ is listed in Table~\ref{TableI}. Here we do not prepare to detailed discuss which model is more suitable for our data, and emphasize that the $\rho(T)$ data for the $x$$\lesssim$0.5 films obey the widely observed temperature dependence of granular hopping conduction, Eq.~(\ref{EqESVRH}), from $\sim$40 down to $\sim$8\,K.

Figure~\ref{LiRTLTLX}~(b) shows the logarithm of the resistivity as a function of the reciprocal of temperature from 10\, down to 2\,K for the $x$$<$$0.50$ films, as indicated. The values of the current applied to the films are also indicated in Fig.~\ref{LiRTLTLX}~(b). Below $\sim$7\,K, the resistivity increases more quickly with decreasing temperature than that in $T$$>$$T_c$ (where $T_c$ is approximately equal to the superconducting transition temperature of bulk Pb), and the logarithm of the resistivity varies linearly with $1/T$ from $\sim$7\,K to $T_s$ ($T_s$ is the temperature below which the linear dependence is not satisfied and the resistivity tends to be saturated, the value of $T_s$ is sample dependent and varies from $\sim$3 to $\sim$4\,K for our films). Below $T_s$, the resistivity gradually approaches saturation with further decreasing temperature. The saturation effect in low temperature regime was also observed in quench-condensed Pb granular films and had been ascribed to the negative electroresistance effect below $T_c$~\cite{no52}. The inset of Fig.~\ref{LiRTLTLX}(b) shows the resistivity as a function of $T^{1/2}$ for the $x$$=$$0.48$ film below $\sim$7\,K. The $\rho(T)$ data clearly deviates from the solid straight line, indicating that the experimental data cannot be described by Eq.~(\ref{EqESVRH}) at $T$$\lesssim$$T_c$.

In the superconductor-insulator granular system, both Cooper pair hopping and single electron hopping processes could occur below $T_c$ at the insulating regime. For $E_c$$>$$\Delta$ (where $\Delta$ is the superconducting gap), the transport is mediated by single electron hopping (or tunneling), while the hopping of Cooper pairs dominates the transport process for the opposite case~\cite{no33, no53}. In the former situation, the temperature dependence of the concentration of single electron excitations in superconducting granules obeys $n\propto\exp(-\Delta/k_BT)$ for $T$$<$$T_c$. Thus the resistivity variation with temperature can be written as~\cite{no15},
\begin{equation}\label{EqActiavtion}
\rho=\rho_{02}\exp\left(\frac{\Delta}{k_B T}\right),
\end{equation}
where $\rho_{02}$ is a prefactor. For the granular superconductor systems, Efetov once gave an effective Hamiltonian and constructed an analytically solvable model, in which Eq.~(\ref{EqActiavtion}) could be naturally obtained~\cite{no54}. Firstly, we assume $\Delta$ to be a constant and designate it by $\Delta_{01}$. The predications of Eq.~(\ref{EqActiavtion}) are least-squares fitted to the experimental $\rho(T)$ data in the $\ln\rho\propto 1/T$ region and shown by the solid straight lines in Fig.~\ref{LiRTLTLX}~(b). The fitted values of the adjustable parameters $\rho_{02}$ and $\Delta_{01}$ are listed in Table~\ref{TableI}. Inspection of Table~\ref{TableI} indicates that values of $\Delta_{01}$ are $\sim$2.40$\times 10^{-22}$\,J (for all the films), which is almost identical to the zero temperature superconducting gap $\Delta_0$ of Pb ($\sim$2.19$\times 10^{-22}$\,J) obtained in single electron tunneling spectra measurements~\cite{no55}. Thus our experimental results indicate that Eq.~(\ref{EqActiavtion}) is quantitatively applicable in the 3D Pb$_x$(SiO$_2$)$_{1-x}$ granular superconductor films in the insulating regime.
%The $\ln(\rho/\rho_{01})\propto T^{-1}$ behavior was also observed in quench-condensed Sn films, ?? films, and ?? films. However, in these previous reports, whether the value of the parameter $\Delta$ is comparable with the local superconducting gap of the metallic granules is still enigmatic.

Now, we consider the influence of temperature on the width of the superconducting gap. We rewrite Eq.~(\ref{EqActiavtion}) as $\ln\rho=\ln\rho_0 + \Delta_0\delta/T$ with $\delta=\Delta(T)/\Delta_0$. The temperature dependence of $\Delta(T)$ can be obtained from the Bardeen-Cooper-Schrieffer (BCS) theory~\cite{no56}. The value of $\Delta_0$ is then obtained form the slope of the $\ln\rho$ versus $\delta/T$ plot. The values of $\Delta_0$ (denoted as $\Delta_{02}$) for the $x$$=$$0.47$, 0.48, 0.49, and 0.50 films are summarized in Table~\ref{TableI}. The magnitudes of $\Delta_{02}$ are about 15\% to 30\% less than that obtained by single electron tunneling method~\cite{no55}. Thus the relative deviation of $\Delta_{02}$ is larger than that by assuming $\Delta$$\equiv$$\Delta_0$ in Eq.~(\ref{EqActiavtion}), which in turn indicates that Eq.~(\ref{EqActiavtion}) can be safely used by treating $\Delta$ as a constant.

Recently, the hopping transport in granular superconductors in the weak-coupling insulating regime has been theoretically investigated~\cite{no33}. It has been
proposed that the inelastic cotunneling mechanism dominates the single electron hopping process in the temperature region $T_1$$\lesssim$$T$$\lesssim$$T_c$, where $T_1\approx 0.1 \sqrt{E_c\delta}$. Due to opening the superconducting gap, the temperature dependence of the resistivity obeys an activation form~\cite{no33},
\begin{equation}\label{EqIneTunneling}
\rho=\rho_{03}\left(\frac{\bar{E}^2}{4g_T\Delta k_BT}\right)^N \exp\left(\frac{2N\Delta}{k_BT}\right),
\end{equation}
where $\bar{g}_T$ and $\bar{E}$$\sim $$E_c$ are the typical values of the conductance and Coulomb correlation energy, respectively, $\rho_{03}$ is a constant, and $N$ is the typical tunneling order. The value of charging energy $E_c$ (or $\bar{E}$) can be estimated through $E_c =e^2/(4\pi\epsilon_0\tilde{\kappa} a)$, where $a$ is the mean grain diameter, $\epsilon_0$ is the permittivity of free space, and $\tilde{\kappa}=\epsilon_r[1+(a/2s)]$ (with $\epsilon_r$ being the dielectric constant of the surrounding medium and $s$ the separation between two adjacent grains) is the effective dielectric constant. Taking $\Delta$ as the gap value of bulk Pb at $T$$=$$0$, and $N$, $\rho_{03}$, and $\bar{g}$ as adjustable parameters, we compare our experimental $\rho(T)$ data with the theoretical predications of Eq.~(\ref{EqIneTunneling}). The results indicate that the  $\rho(T)$ data can be well described by Eq.~(\ref{EqIneTunneling}) from $\sim$7\, down to $\sim$3\,K for the $x$$\simeq$0.47, 0.48, 0.49, 0.50 films. However, the values of the adjustable parameter $N$ are $N\simeq0.50$ for all the films. Theoretically, the tunneling order $N$ should be $N\gtrsim 2$. Thus the fact $N\simeq0.50$ for all films reveals that the inelastic cotunneling process does not occur in the Pb$_x$(SiO$_2$)$_{1-x}$ films in insulating regime. On the other hand, the temperature behavior of $\rho(T)$ in Eq.~(\ref{EqIneTunneling}) is mainly determined by the exponential factor, thus $N\simeq0.50$ implies that the model described by Eq.~(\ref{EqActiavtion}) can be well account for the hopping transport in the 3D Pb$_x$(SiO$_2$)$_{1-x}$ films.

\begin{figure}
\begin{center}
\includegraphics[scale=1]{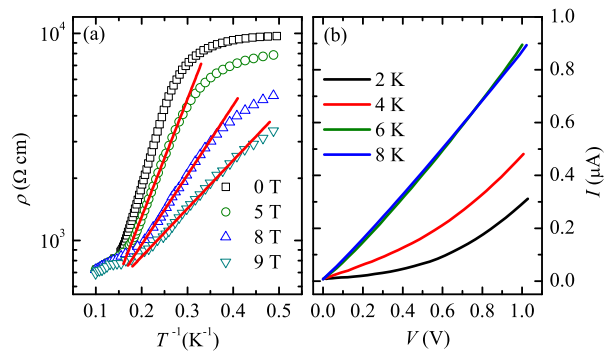}
\caption{(a)  Logarithm of resistivity as a function of the reciprocal of temperature from $\sim$10 down to 2\,K under different fields for the $x$$\simeq$$0.48$ film. (b) Current $I$ versus voltage $V$ at different temperatures for the $x$$\simeq$$0.5$ film.}\label{FigRTHandIV}
\end{center}
\end{figure}

Figure~\ref{FigRTHandIV}(a) shows the resistivity as a function of  the reciprocal of temperature under different fields from 10 to 2\,K for the $x$$\simeq$$0.48$ film. At a certain temperature, the resistivity decreases with increasing field, which becomes remarkable below $\sim$7\,K. Since the magnetic field could lead to suppression of the superconductor ordering parameter and decrease of the Josephson coupling, the negative magnetoresistance clearly indicates that the hopping transport in the $x$$\lesssim$0.5 films is governed by the hopping of single electron instead of Cooper pairs. When the magnetic field is applied, the $\ln(\rho/\rho_{02})\propto 1/T$ law is still satisfied below $\sim$7\,K. The $\rho$-$T$ curves under 5, 8, and 9\,T are also least-squares fitted to Eq.~(\ref{EqActiavtion}) (solid straight lines), and the corresponding values of $\Delta$ are 1.81$\times$10$^{-22}$, 1.07$\times$10$^{-22}$, and 7.39$\times$10$^{-23}$\,J, respectively. The field suppresses the width of the local superconducting gap, however, local superconductivity has not been completely suppressed even the magnitude of the field is as large as 9\,T.

In Fig.~\ref{FigRTHandIV}(b), we present the current $I$ as a function of voltage $V$ for the $x$$\simeq$$0.5$ film. Below $\sim$7\,K, the current varies nonlinearly with the voltage at higher voltage region and no hysteresis is observed in the $I$-$V$ loci. At a certain temperature, the resistance ($V/I$) decreases with increasing the applied voltage at higher voltage region, which indicates a negative electroresistance effect in the film. This confirms that the saturation effect of resistivity at low temperature region originates from the negative electroresistance effect of the films~\cite{no52}. In the $\rho$-$T$ data measurements, although we take the current as low as possible, as indicated in Fig.~\ref{LiRTLTLX}(b), the saturation effect is still not avoided.

\begin{figure}
\begin{center}
\includegraphics[scale=0.95]{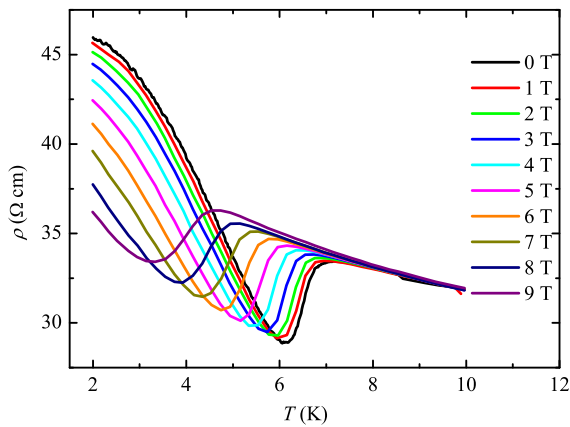}
\caption{Resistivity versus temperature under different field for the $x$$=$$0.52$ film.}\label{FigRTReentrant}
\end{center}
\end{figure}

\subsection{Reentrant behavior of the resistivity}
Figure~\ref{FigRTReentrant} shows the resistivity as a function of temperature from 10 to 2\,K for the $x$$\simeq$0.52 film at different field, as indicated. In fact, the resistivities variation with temperature and magnetic field for the 0.50$<$$x$$<$$0.57$ films are similar to those of the  $x$$\simeq$0.52 film, respectively. From Fig.~\ref{FigRTReentrant}, one can see that the temperature coefficient of resistivity keep negative above $\sim$7.2\,K, then the resistivity sharply decreases with decreasing temperature, reaches its minimum at $T_{\mathrm{min}}$, and then increases again with further decreasing temperature. This reentrant  behavior of resistivity is prevalent in the 2D granular superconductor films near the SI transition~\cite{no7} and absent in disordered homogeneous superconductor films~\cite{no15, no31, no57}. Kunchur \emph{et al} observed the reentrant behavior in 3D granular Al films~\cite{no30}. Combing with the results of Kunchur \emph{et al}, it can be concluded that the reentrant behavior of resistivity is also prevailing in 3D granular superconductors and occurs in the insulating side of the films with $x$ slightly lower than $x_c$.

Fisher once gave an explanation to the reentrant behavior of the resistivity~\cite{no58}. According to Fisher, the drop in resistivity just below $T_c$ (where $T_c$ is the superconducting transition temperature of the superconductor granules) arises from the shorting out of a portion of the sample by superconducting regions, and quantum tunneling will induce phase slips across the links (barriers), leading to an enhancement of the resistance at sufficiently low temperature. In framework of this scenario, the magnetoresistance at low temperature regime would be positive since the magnetic field results in suppression of the order parameter of superconductor. Form Fig.~\ref{FigRTReentrant}, one can see the magnetoresistance of the $x$$\simeq$0.52 films is negative at low temperature and its absolute value is $\sim$21\% at 9\,T and 2\,K. This indicates that single electron hopping (or tunneling) could also be the dominated transport mechanism at low temperature for the films with reentrant resistivity behavior. In $T_\mathrm{min}$$<$$T$$<$$T_c$ region, the magnetoresistance is positive and the temperature behavior of resistivity under magnetic field is similar to that of the film with global superconductivity in temperature slighly below $T_c$. Hence below $T_c$ Cooper pairs are formed in the Pb particles and the portions formed by directly connected Pb particles or Pb particles with strong Josephson coupling would be shorted, which results in a sharp drop in resistivity. Below $T_\mathrm{min}$, the single electron hopping or tunneling will dominate the transport process, and the reduction of electron concentration leads to the enhancement of resistivity with further decreasing temperature.

\subsection{Dissipation effect in the metallic regime}
\begin{figure}
\begin{center}
\includegraphics[scale=1]{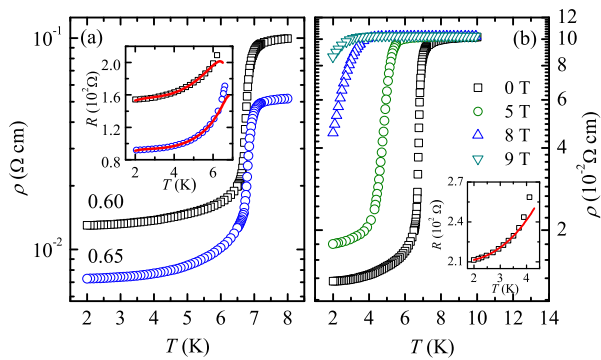}
\caption{(a) Resistivity as a function of temperature for $x$$=$$0.60$ and 0.65 films. Inset: resistance versus temperature below $\sim$6.5\,K for the two films. The solid curves are fits to Eq.~(\ref{EqTAPSandMQT}). (b) Resistivity as a function of temperature under different fields for the $x$$=$$0.60$ film. Inset: resistance versus temperature under 5\,T below $\sim$4.5\,K for the film. The solid curve is the fit to Eq.~(\ref{EqTAPSandMQT}). }\label{FigRTDissipation}
\end{center}
\end{figure}

Figure~\ref{FigRTDissipation} (a) shows the resistivity as a function of temperature for the $x$$\simeq$$0.60$ and 0.65 films, as indicated. The electrical transport properties of the $0.60$$\lesssim$$x$$\lesssim$0.72 films are similar to that of the two representative films. Surprisingly, although the $x$$\simeq$$0.60$ and 0.65 films reveal metallic behavior above $T_c$, i.e., the temperature coefficients of the resistivity $(1/\rho) \mathrm{d}\rho/\mathrm{d}T$ is positive, global superconductivity is not appear at any film down to 2\,K. For the two films, the resistivities drop sharply below $\sim$7\,K and are reduced to $\sim$$17\%$ ($x$$\simeq$$0.60$ film) and $\sim$$20\%$ ($x$$\simeq$$0.65$ film) of the value of the normal state at $\sim$6\,K, respectively. Then the resistivities gradually decrease with further decreasing temperature. Global superconductivity is present in the $x$$\gtrsim $$0.73$ films. Here we focus our discussion on the $0.60$$\lesssim$$x$$\lesssim$0.72 films.

The dissipation effect in superconductor has been observed in quasi 1D superconducting wires~\cite{no34, no35, no36, no37, no38, no39} and 2D superconducting films~\cite{no10,no17} and has been intensively investigated both in experimental and  theoretical sides. The earlier understanding of dissipation in 1D superconductors at $T$$\lesssim$$T_c$ is based on the work of Langer and Ambegaokar~\cite{no59}, and McCumber and Halperin~\cite{no60}. According to their picture, the current-carrier state is only metastable, and dissipation occurs when the system passes over a free energy barrier $\Delta F$ to a lower energy state via thermal activation. This process, being called thermal activation phase slip (TAPS), could result in a resistance with scale $\exp (-\Delta F/k_B T)$. From 1988 to 1991,  Giordano investigated the transport properties of small-diameter In and Pb-In wires, and has found the thermal activation phase slip process alone cannot explain the dissipation effect well below $T_c$~\cite{no34, no35, no36}. He has suggested that the phase slippage caused by the quantum-mechanical tunneling of the order parameter through the free-energy barrier plays dominant role at low temperatures.
This mechanism for phase slippage is also called macroscopic quantum tunneling (MQT), since it is analogous to the MQT occurs in tunneling junction, superconducting quantum-interference device, and other systems~\cite{no61, no62, no63}.

According to the percolation theory~\cite{bookRichard}, above $x_c$ there is at least one conducting path, in which each metallic particle geometrically connects with the nearest neighbors. Thus each conducting path in the metal-insulator nanogranular systems with $x$ slightly greater than $x_c$ can be reasonably treat as a metal nanowire. We then analyze the resistance tails below $T_c$ for the $0.60$$\lesssim$$x$$\lesssim$0.72 films by considering the combination effect of TAPS and MQT. In low current limit, the resistance involving both TAPS and MQT effects can be written as~\cite{no34, no35, no36, no37}
\begin{eqnarray}\label{EqTAPSandMQT}
R & = &R_\mathrm{TAPS} + R_\mathrm{MQT} \nonumber\\
  & = &c_1 \frac{\pi \hbar^2 \Omega_{\mathrm{TAPS}}}{2e^2 k_B T } e^{-\alpha\Delta F/k_B T} + c_2\frac{\pi \hbar^2 \Omega_\mathrm{MQT}}{2e^2 (\hbar/\tau_\mathrm{GL})}e^ {-\beta\Delta F\tau_\mathrm{GL}/\hbar} \nonumber\\
  &   &
\end{eqnarray}
where $c_1$, $c_2$, $\alpha$, and $\beta$ are possible numerical factors, $\Omega_\mathrm{TAP}=(L/\xi)(\Delta F/k_B T)^{1/2}(1/\tau_\mathrm{GL})$ and $\Omega_\mathrm{MQT}=(L/\xi)[\Delta F/(\hbar/\tau_\mathrm{GL})]^{1/2}(1/\tau_\mathrm{GL})$ are the attempt frequencies of thermal activation phase slip and quantum phase slip, respectively, and $\Delta F= (8\sqrt{2}/3)(H_c^2/8\pi)A\xi$ is the magnitude of the free-energy barrier. In these expressions, $L$ is the length of the conducting path, $H_c$ and $\xi$ are the thermodynamic critical field and coherence length of the material, $A$ is the cross-section area of the conducting path, and $\tau_\mathrm{GL}=\pi\hbar/8k_B(T_c-T)$ is the characteristic relaxation time in the time-dependent Ginzburg-Landau theory.

The thermodynamic critical field and coherence length of bulk Pb are 0.08\,T and 87\,nm~\cite{no64}, respectively. Taking the length $L$ and area of the cross section $A$ of each conducting path as $L\simeq5$\,mm and $A \simeq a^2$, respectively, we compare the low temperature resistance $R(T)$ data of the $x$$\simeq $$0.60$ and 0.65 films with the theoretical predication of Eq.~(\ref{EqTAPSandMQT}). The inset of Fig.~\ref{FigRTDissipation}(a) shows the enlarged image of temperature dependence of resistance from 2 to $\sim$6\,K for the $x$$\simeq$$0.60$ and 0.65 films. The solid curves are fits to Eq.~(\ref{EqTAPSandMQT}). Clearly, the experimental $R(T)$ data from 2 to $\sim$6\,K can be well described by Eq.~(\ref{EqTAPSandMQT}). The values of the adjustable parameters are $c_1\simeq 5.4 \times 10^{-6}$, $c_2\simeq 5.5 \times 10^{-5}$, $\alpha\simeq 6976$, and $\beta \simeq 1944$ for the $x$$\simeq$$0.60$ film and $c_1\simeq 3.3 \times 10^{-6}$, $c_2\simeq 2.5 \times 10^{-5}$, $\alpha\simeq 5124$, and $\beta \simeq 648$ for the $x$$\simeq$$0.65$ film. The parameters $c_1$ and $c_2$ are both far less than unity, while $\alpha$ and $\beta$ much greater than unity.  These deviations should not be taken too seriously since both the theories of TAPS and MQT are extremely qualitative~\cite{no35}. On the other hand, there are more than one conducting paths in the $x$$\gtrsim$$x_c$ films. Hence all the conducting paths connected in parallel and resistance of the film is about $R_{\mathrm{sig}}/\frak{n}$, where $R_{\mathrm{sig}}$ is the resistance of a single conducting path and $\frak{n}$  is the number of conducting paths between the two voltage electrodes. Thus the value of $c_1$ ($c_2$) should be $1/\frak{n}$ of that for a single conducting path. In addition, the values of the $L$ and $A$ are also quite qualitative, which seriously affect the magnitudes of the numerical factors $c_1$, $c_2$, $\alpha$, and $\beta$.

Figure~\ref{FigRTDissipation}(b) shows the temperature dependence of the resistivity for the $x$$=$$0.60$ film at different fields. Inspection of Fig.~\ref{FigRTDissipation}(b) indicates that the transition temperature decreases with increasing magnetic field and the magnetoresistance is positive below $\sim$7\,K~\cite{no65}. Even a field with magnitude of 5\,T is applied, the variation trend of the $\rho(T)$ curve is retained, i.e. the resistance tail also exists at low temperature. In addition, the $R(T)$ curve at 5\,T can be also described by Eq.~(\ref{EqTAPSandMQT}) [solid curve in the inset of Fig.~\ref{FigRTDissipation}(b)] below $\sim$4\,K, which indicates that both the TAP and MQT processes are also the main mechanisms of the dissipation effect. The values of the adjustable parameters are $c_1\simeq 1.6 \times 10^{-6}$, $c_2\simeq 8.3 \times 10^{-5}$, $\alpha\simeq 3629$, and $\beta \simeq 887$. We note in passing that the dissipation effect in our films is not caused by the
granular effect of the system. The positive TCR from 300 down to $\sim$10\,K has indicated that the nearest neighbor Pb granules in the conducting path have geometrically connected and there is no insulator barrier between them.

\section{Conclusion}
We have investigated the electrical transport properties of Pb$_x$(SiO$_2$)$_{1-x}$ nanogranular films with $0.47$$\lesssim$$x$$\lesssim$0.72. The percolation threshold of this system lies between 0.57 and 0.60. For the 0.47$\lesssim$$x$$\lesssim$0.50 films, the temperature dependence of resistivities obeys $\rho=\rho_{02}\exp(\Delta/k_BT)$ law from $\sim$7\,K to $T_s$, and the experimental values of $\Delta$ are comparable to that obtained in single electron tunneling spectra measurement. Below $T_c$ the magnetoresistance is negative and its absolute value is much larger than that above $T_c$, which indicates the single electron hopping or tunneling dominates the transport process below $T_c$. For the 0.50$<$$x$$<$0.57 films, the temperature dependence of resistivities reveals reentrant behavior below $T_c$. We argue that the drop in resistivity just below $T_c$ arises from the shorting out of a portion of the sample by superconducting regions, the single electron hopping or tunneling governs the low-temperature transport process and the reduction of the single electron concentration with decreasing temperature leads to an enhancement of the resistivity at sufficiently low temperature. Although the TCRs of the 0.60$\lesssim$$x$$\lesssim$0.72 films are all positive form 300 down to 10\,K, global superconductivity is not present in these films. The resistivities of these films decrease sharply with decreasing temperature at temperature slightly below $T_c$, and then decrease slowly with further decreasing temperature. We treat the conducting paths as Pb nanowires and have found that the $R(T)$ data below $T_c$ can be well described by the model that includes both thermally activated phase slips and quantum phase slips.

\begin{acknowledgments}
This work is supported by the National Natural Science Foundation of China through Grant No. 11774253.
\end{acknowledgments}


\begin{thebibliography}{00}
\bibitem{no65} Although global superconductivity is not present in this film, we also designate a transition temperature at which the resistivity has dropped to $\rho_N/2$ with $\rho_N$ being the normal state resistivity at 10\,K.
\end{thebibliography}
\end{document}